\documentclass[a4paper]{article}
\usepackage[utf8x]{inputenc}
\usepackage[left=3cm,right=3cm,top=3cm,bottom=3cm]{geometry}
\usepackage[linktoc=all,hidelinks]{hyperref}
\usepackage{cite}

\usepackage{amsmath}
\usepackage{amsfonts}
\usepackage{amssymb}

\usepackage{tensor}
\usepackage{slashed}

\usepackage{bbm}

\usepackage{graphicx}
\usepackage{young}
\usepackage[vcentermath]{youngtab}

\usepackage{color}

\definecolor{darkgreen}{rgb}{0.2,0.7,0.2}

\newcommand{\pd}{\partial}
\newcommand{\wt}{\widetilde}

\newcommand{\al}{\alpha}

\newcommand{\si}{\sigma}

\newcommand{\be}{\begin{equation}}
\newcommand{\ee}{\end{equation}}
\newcommand{\bea}{\begin{eqnarray}}
\newcommand{\eea}{\end{eqnarray}}

\numberwithin{equation}{section}

\begin{document}

%%%%%%%%%%%%%%%%%%%%%%%%%%%%%%%%%%%%%%%%%%%%%%%%%%
% Title page & table of contents
\thispagestyle{empty}
\begin{center}

\vspace*{50pt}
{\LARGE \bf A note on the double dual graviton}

\vspace{30pt}
{Marc Henneaux${}^{\, a,b}$, Victor Lekeu${}^{\, a,c}$ and Amaury Leonard${}^{\, d}$}

\vspace{10pt}
\texttt{henneaux@ulb.ac.be, vlekeu@ulb.ac.be, amaury.leonard@aei.mpg.de}

\vspace{20pt}
\begin{enumerate}
\item[${}^a$] {\sl \small
Universit\'e Libre de Bruxelles and International Solvay Institutes,\\
ULB-Campus Plaine CP231, B-1050 Brussels, Belgium}
\item[${}^b$] {\sl \small
Coll\`ege de France, 11 place Marcelin Berthelot, 75005 Paris, France\\}
\item[${}^c$] {\sl \small
The Blackett Laboratory, Imperial College London,\\Prince Consort Road London SW7 2AZ, U.K.}
\item[${}^d$] {\sl \small
Max-Planck-Institut f\"{u}r Gravitationsphysik (Albert-Einstein-Institut),\\
Am M\"{u}hlenberg 1, DE-14476 Potsdam, Germany}
\end{enumerate}

\vspace{50pt}
{\bf Abstract} 
\end{center}

\noindent
The (free) graviton admits, in addition to the standard Pauli-Fierz description by means of a rank-two symmetric tensor, a description in which one dualizes the corresponding $(2,2)$-curvature tensor on one column to get a $(D-2,2)$-tensor, where $D$ is the spacetime dimension.  This tensor derives from a gauge field with mixed Yound symmetry $(D-3,1)$ called the ``dual graviton" field.   The dual graviton field is related non-locally to the Pauli-Fierz field (even on-shell), in much the same way as a $p$-form potential and its dual $(D-p-2)$-form potential  are related in the theory of an abelian $p$-form.  Since the Pauli-Fierz field has a Young tableau with two columns (of one box each), one can contemplate a double dual description in which one dualizes on both columns and not just on one.  The double dual curvature is now a $(D-2,D-2)$-tensor and derives from a gauge field with $(D-3, D-3)$ mixed Young symmetry, the ``double dual graviton" field.  We show, however, that the double dual graviton field is algebraically and locally related to the original Pauli-Fierz field and, so, does not provide a truly new description of the graviton. From this point of view, it plays a very different role from the dual graviton field obtained through a single dualization. We also show that these equations can be obtained from a variational principle in which the variables to be varied in the action are (all) the components of the double-dual field as well as an auxiliary field with $(2,1)$ Young symmetry.  By gauge fixing the shift symmetries of this action principle, one recovers the Pauli-Fierz action. Our approach differs from the interesting approach based on parent actions and covers only the free, sourceless theory. Similar results are argued to hold for higher spin gauge fields.  

\newpage

\setcounter{tocdepth}{2}
\tableofcontents

\newpage

%%%%%%%%%%%%%%%%%%%%%%%%%%%%%%%%%%%%%%%%%%%%%%%%%%
\section{Introduction}
\setcounter{equation}{0}

This paper is dedicated to Peter Freund, who had precient intuitions about the importance of dual formulations of  field theories and the role played by fields with mixed Young tableau symmetry in that context \cite{Curtright:1980yj}.    

It is well known that Abelian $p$-form gauge fields $A^{(p)}$ admit two dual descriptions.  The first one is based on the $p$-form potential $A^{(p)}$ itself, from which one derives the curvature $(p+1)$-form $F^{(p+1)} = d A^{(p)}$, which fulfills 
\be
d F^{(p+1)} = 0 \label{Bianchi0}
\ee
(identically).  The Maxwell equations of motion are 
\be
d H^{(D-1-p)} = 0 \label{EOM0}
\ee
where
$H^{(D-1-p)}$ -- a $(D-p-1)$-form -- is the Hodge dual of $F^{(p+1)}$,
\be
H^{(D-1-p)} = \; ^*\!F^{(p+1)}.
\ee

Because of (\ref{EOM0}), one can introduce a dual $(D-p- 2)$-form potential $B^{(D-p-2)}$ such that $H^{(D-1-p)} = d B^{(D-p-2)}$.  The equations of motion for $B^{(D-p-2)}$ are (\ref{Bianchi0}), which reads $d \; ^*\!H^{(D-1-p)} = 0$.  Bianchi identities and equations of motion are exchanged as one goes from one description to its dual description.  Note that the relation between the $p$-form potential $A^{(p)}$ and its dual $B^{(D-p-2)}$ is non-local. If one were to introduce sources, electric (respectively magnetic) sources for $A^{(p)}$ would appear as magnetic (respectively electric) sources for $B^{(D-p-2)}$.  One sometimes speaks of electric-magnetic duality for this reason.

In terms of Young tableaux, $A^{(p)}$ is described by a Young tableau with a single column with $p$ boxes and its curvature $F^{(p+1)}$ is described by a Young tableau with a single column with $p+1$ boxes.  The dual curvature $H^{(D-1-p)}$ is obtained by dualizing on the only column there is, and is described by a Young tableau with a single column with $D- p-1$ boxes.   Finally, the dual potential $B^{(D-p-2)}$ is described by a Young tableau with a single column with $D- p-2$ boxes.

Gravitational duality defined in terms of Hodge duality operations on the curvature tensor was considered a while ago in the papers \cite{Hull:2000zn,Hull:2001iu} and further studied in \cite{deMedeiros:2002qpr}.  Equivalent definitions involving the connection were independently given in \cite{West:2001as}. Since the Pauli-Fierz field has a Young tableau $\yng(2)$ with two columns and hence a curvature tensor that is described by a Young tableau $\yng(2,2)$ also with two columns, one can now consider two different types of duality.  One can dualize on a single column\footnote{Which column one takes is of course a matter of choice, since there is symmetry between the columns.} or one can dualize on the two different columns, leading respectively to the ``dual graviton" and the ``double dual graviton".  The dual graviton was effectively considered earlier in \cite{Curtright:1980yk} and more recently in \cite{West:2001as,Damour:2002cu} in connection with hidden symmetries of gravity.

The purpose of this note is to show that contrary to the dual graviton field that is non-locally related to the Pauli-Fierz field,  the double dual graviton field can in fact be viewed as a mere algebraic rewriting of the Pauli-Fierz field.  From that point of view the double dual graviton field does not really bring a truly new description of a massless spin-2 particle.  We also argue that the same property holds for higher spin gauge fields, for which duality was defined in terms of curvatures in \cite{Hull:2001iu,Bekaert:2003az}.  Only the single dual is a truly new field. Our results, relevant to the covariant Lagrangian formulation, are in line with the light cone gauge considerations of \cite{Francia:2005bv}.

\section{Gravitational duality in five spacetime dimensions}
\label{FiveD}

We start by reviewing the various dual descriptions of the graviton in the case $D=5$, which illustrates the main point.

%\subsection{Duality equations}

%\subsubsection{Pauli-Fierz field}
\subsection{Pauli-Fierz field}

The standard description of a free massless spin-two particle involves a symmetric tensor $h_{\mu \nu} = h_{\nu \mu}$ (Young symmetry type $(1,1) \equiv \yng(2)$), the ``Pauli-Fierz field".  This field is invariant under the gauge symmetries
\be
\delta h_{\mu \nu} = 2\partial_{(\mu} \xi_{\nu)} \label{LinDiff}
\ee
(``linearized diffeomorphisms").

A complete set of invariants under (\ref{LinDiff}) is given by the ``Riemann" (or ``curvature") tensor $R_{\lambda \mu \rho \sigma}$ defined by
\begin{equation}
R_{\lambda \mu \rho \sigma} = -\frac{1}{2} \left(\partial_\lambda \partial_\rho h_{\mu \sigma} -  \partial_\mu \partial_\rho h_{\lambda \sigma} - \partial_\lambda \partial_\sigma h_{\mu \rho} + \partial_\mu \partial_\sigma h_{\lambda \rho}\right). \label{Riemann0}
\end{equation}
The Riemann tensor is of Young symmetry type
$$
(2,2) \equiv \yng(2,2)
$$
i.e. fulfills the algebraic identities 
\begin{equation}
R_{\lambda \mu \rho \sigma} =  R_{[ \lambda \mu ]\rho \sigma}, \; \; \; \; R_{\lambda \mu \rho \sigma} = R_{\lambda \mu [\rho \sigma ]}, \; \; \; \; \; R_{[\lambda \mu \rho] \sigma} = 0. \label{AlgebraicI}
\end{equation}
The Riemann tensor also fulfills the differential Bianchi identity
\begin{equation}
\partial_{[\alpha_1} R_{\alpha_2 \alpha_2] \beta_1 \beta_2} = 0  \label{BianchiI}
\end{equation}
Conversely, given a tensor $R_{\lambda \mu \rho \sigma}$ fulfilling the conditions (\ref{AlgebraicI}) and (\ref{BianchiI}), there is a tensor $h_{\lambda \mu}$ from which $R_{\lambda \mu \rho \sigma}$ derives as in (\ref{Riemann0}).  The tensor $h_{\lambda \mu}$ is determined up to the gauge transformations (\ref{LinDiff}).

The linearized Einstein equation are
\begin{equation}
R_{\lambda \rho} = 0, \label{LinearizedEinstein}
\end{equation}
where $R_{\lambda \rho}$ is the linearized Ricci tensor,
\begin{equation}
R_{\lambda \rho} = R_{\lambda \mu \rho \sigma} \eta^{\mu \sigma} . \label{Ricci0}
\end{equation}

%\subsubsection{The Curtright field or ``dual" graviton}
\subsection{The Curtright field or ``dual" graviton}
\label{CurtrightField}

If one dualizes the Riemann tensor on its first column, one gets a tensor $E_{\alpha_1 \alpha_2 \alpha_3 \beta_1 \beta_2}$,
\be 
E_{\alpha_1 \alpha_2 \alpha_3 \beta_1 \beta_2} = \frac{1}{3!} \epsilon_{\alpha_1 \alpha_2 \alpha_3 \lambda_1 \lambda_2} R^{\lambda_1 \lambda_2}_{\; \; \; \; \; \; \; \; \beta_1 \beta_2}
\ee
which is traceless on account of the cyclic identity $ R_{[\lambda_1 \lambda_2 \beta_1] \beta_2} = 0$ for the Riemann tensor.   The equation of motion (\ref{LinearizedEinstein}) implies moreover that $E_{\alpha_1 \alpha_2 \alpha_3 \beta_1 \beta_2}$ has the $(3,2)$ Young symmetry,
$$\yng(2,2,1).$$
Furthermore, $E_{\alpha_1 \alpha_2 \alpha_3 \beta_1 \beta_2}$ fulfills the necessary Bianchi identities that guarantee that it can be written as 
\be
E_{\alpha_1 \alpha_2 \alpha_3 \beta_1 \beta_2} = \partial_{[\alpha_1} T_{ \alpha_2 \alpha_3][ \beta_1, \beta_2]}
\ee
for some field $T_{ \alpha_2 \alpha_3 \beta_1}$ that has the $(2,1)$ Young symmetry \cite{Coho,DuboisViolette:1999rd,DuboisViolette:2001jk,Bekaert:2002dt},
$$
\yng(2,1),
$$
 i.e.
\begin{equation}
T_{\alpha_1 \alpha_2 \beta}= -T_{\alpha_2 \alpha_1 \beta}, \; \; \; \; T_{[\alpha_1 \alpha_2 \beta]} = 0.
\end{equation}
This field $ T_{\alpha_1 \alpha_2 \beta}$ is determined from its ``curvature" $E_{\alpha_1 \alpha_2 \alpha_3 \beta_1 \beta_2}$ up to the gauge transformations
\begin{equation}
\delta T_{\alpha_1 \alpha_2 \beta} = 2 \partial_{[\alpha_1} \si_{\alpha_2] \beta} + 2 \partial_{[\alpha_1} \al_{\alpha_2] \beta} - 2 \partial_{\beta} \al_{\alpha_1 \alpha_2}
\end{equation}
where $\si_{\alpha \beta}$ and $\al_{\alpha \beta}$ are symmetric and antisymmetric tensor fields, respectively,
\begin{equation}
\si_{\alpha \beta} = \si_{\beta \alpha}, \; \; \; \; \al_{\alpha \beta} = - \al_{\beta \alpha}.
\end{equation}

The equations of motion for the $T$-field are that the trace of its curvature vanishes,
\begin{equation}
E_{\alpha_1 \alpha_2 \alpha_3 \beta_1 \beta_2} \eta^{\alpha_3 \beta_2} = 0. \label{EOMTField}
\end{equation}
When these equations of motion hold, one can go backwards and recover the Pauli-Fierz field $h_{\mu \nu}$, which obeys the linearized Einstein equations.  It is easy to verify that the relationship between $h_{\mu \nu}$ and $T_{\alpha_1 \alpha_2 \beta}$ is non-local (involves spacetime integrations), just as the relationship between a $p$-form and its $(n-p-2)$-form dual is.

In \cite{Curtright:1980yk}, the theory of generalized gauge fields described by higher rank tensors which are neither completely symmetric nor completely antisymmetric, and which obey equations of motion of the type (\ref{EOMTField})  was initiated.   
In particular, the case of a $(2,1)$-tensor was investigated in depth. For that reason, one sometimes calls a gauge field with the $\yng(2,1)$-Young symmetry the ``Curtright field" (see also \cite{Aulakh:1986cb,Labastida:1986gy,Labastida:1987kw}).  Alternatively, because of its relation with the graviton, one also uses the terminology ``dual graviton", or even ``single-dual graviton" to emphasize that dualization of the curvature is performed on a single column only.  More information on the $(2,1)$-tensor gauge field can be found in  \cite{Hull:2001iu,deMedeiros:2002qpr} and \cite{Bekaert:2002uh,Bunster:2013oaa}.

%\subsubsection{The double dual}
\subsection{The double dual}

One can also dualize the curvature on both columns and define
\be
M_{\alpha_1 \alpha_2 \alpha_3 \beta_1 \beta_2 \beta_3} = \left(\frac{1}{3!} \epsilon_{\alpha_1 \alpha_2 \alpha_3 \lambda_1 \lambda_2} \right) \left(  \frac{1}{3!} \epsilon_{\beta_1 \beta_2 \beta_3 \mu_1 \mu_2}\right) R^{\lambda_1 \lambda_2 \mu_1 \mu_2} \label{eq:DoubleDualCurv}
\ee
In that way, one gets a tensor that has the $(3,3)$ mixed Young symmetry, with tableau
$$\yng(2,2,2).$$
Using the Bianchi identity and the field equations for $h_{\mu \nu}$, it  can be easily checked that the tensor $M_{\alpha_1 \alpha_2 \alpha_3 \beta_1 \beta_2 \beta_3}$ fulfills  the Bianchi identity that guarantees the existence of a $(2,2)$-field $C_{\alpha_1 \alpha_2  \beta_1 \beta_2}$ such that 
\be
M_{\alpha_1 \alpha_2 \alpha_3 \beta_1 \beta_2 \beta_3} = \partial_{[\alpha_1} C_{\alpha_2 \alpha_3]  [\beta_1 \beta_2 , \beta_3]}
\ee
(see \cite{DuboisViolette:1999rd,DuboisViolette:2001jk}). This $(2,2)$-field is the ``double dual graviton" field.

The equations of motion for the double dual graviton are that the double trace of its ``Riemann tensor'' $M_{\alpha_1 \alpha_2 \alpha_3 \beta_1 \beta_2 \beta_3}$ vanishes  \cite{Hull:2001iu,deMedeiros:2002qpr}, i.e. 
\begin{equation}
M_{\alpha_1 \alpha_2 \alpha_3 \beta_1 \beta_2 \beta_3} \eta^{\alpha_2 \beta_2} \eta^{\alpha_3 \beta_3} = 0. \label{EOMCField}
\end{equation}
This can be seen by direct computation from  (\ref{eq:DoubleDualCurv}), which implies that the ``Einstein 
tensor'' $G_{\alpha \beta \lambda \mu}[C]$ of $C_{\alpha \beta \lambda \mu}$, as defined in \cite{Henneaux:2016opm}, is related to the Riemann tensor of $h_{\mu \nu}$ as
\be
G_{\alpha \beta \lambda \mu}[C]= \frac{1}{18} R_{\alpha \beta \lambda \mu}[h]
\ee
The vanishing of the trace of $R_{\alpha \beta \lambda \mu}[h]$ is equivalent to the vanishing of the trace of $G_{\alpha \beta \lambda \mu}[C]$, i.e. to the vanishing of the double trace of $M_{\alpha_1 \alpha_2 \alpha_3 \beta_1 \beta_2 \beta_3}$.

In fact, spacetime dimension 5 is the ``critical dimension" where the ``Weyl tensor'' of a $(2,2)$-field -- i.e. the trace-free part of its curvature -- identically vanishes.  The Einstein tensor $G_{\alpha \beta \lambda \mu}[C]$ of $C_{\alpha \beta \lambda \mu}$ contains therefore the complete information on the curvature.  If it were to vanish, one would find that the curvature $M_{\alpha_1 \alpha_2 \alpha_3 \beta_1 \beta_2 \beta_3}$ itself should vanish, which is too strong as it would imply, in turn, that the curvature $R_{\alpha \beta \lambda \mu}[h]$ of the Pauli-Fierz field should also vanish. The correct equations are that the double trace (and not the single trace) of the curvature $M_{\alpha_1 \alpha_2 \alpha_3 \beta_1 \beta_2 \beta_3}$ is zero.

To summarize, the graviton has three dual descriptions: one in term of a field $h_{\mu \nu}$ transforming as 
$$\yng(2)\, ,$$
one in term of a field $T_{\alpha \beta \lambda}$ transforming as
$$\yng(2,1)\, ,$$
and one in term of a field $C_{\alpha \beta \lambda \mu}$ transforming as
$$ \yng(2,2)\, .$$  
We shall show, however, that the latter description is not truly new because the field  $C_{\alpha \beta \lambda \mu}$ can be algebraically related to the original field $h_{\mu \nu}$, and  its action can be obtained from the Pauli-Fierz action through algebraic changes of variables.  It is in that sense a straightforward reformulation of the original Pauli-Fierz formulation.

%\subsection{The double dual is not an algebraically independent field}
\section{The double dual is not an algebraically independent field}

%\subsubsection{Relationship between $C$ and $h$}
\subsection{Relationship between $C$ and $h$}
It follows from the definition of the double dual curvature and simple $\epsilon$ algebra that 
\be
M\indices{^{\alpha_1 \alpha_2 \alpha_3}_{\beta_1 \beta_2 \beta_3}} = - \, \delta^{[\alpha_1}_{[\beta_1} R\indices{^{\alpha_2 \alpha_3]}_{\beta_2 \beta_3]}} \label{MR} .
\ee
In \eqref{MR}, we have used the equation of motion $R_{\alpha \beta} = 0$.  

Writing the curvatures in terms of two derivatives of the relevant fields, one has from \eqref{MR}
\be
\partial^{[\alpha_1} \partial_{[\beta_1} \left( C\indices{^{\alpha_2 \alpha_3]}_{\beta_2 \beta_3]}} + \delta^{\alpha_2}_{\beta_2} h\indices{^{\alpha_3]} _{\beta_3]}} \right) = 0\, ,
\ee
an equation that takes the form $D^2 (C + h \delta) = 0$ in terms of the operator $D$ of  \cite{DuboisViolette:1999rd,DuboisViolette:2001jk} that fulfills $D^3=0$.  The corresponding Poincar\'e lemma implies then that $C + h \delta  = D \xi$ for some $(2,1)$-field $\xi$, i.e., writing indices, that
 the dual graviton $C_{\alpha\beta\mu\nu}$ is algebraically related to the Pauli-Fierz field $h_{\mu\nu}$ up to a physically irrelevant gauge transformation,
\be
C_{\alpha \beta \mu \nu} = {\mathbb P}_{(2,2)}\left(\partial_{\alpha} \xi_{\mu\nu \beta} \right)  - \frac{1}{2} \left(\eta_{\mu [\alpha} h_{\beta] \nu} - \eta_{\nu [\alpha} h_{\beta] \mu} \right) \, .\label{Triviality2}
\ee
Equation (\ref{Triviality2}) shows that the double-dual graviton field is ``conformally flat'', i.e. the sum of a $(2,2)$-diffeomorphism parametrized by $\xi$ and a $(2,2)$-Weyl rescaling parametrized by $h$.
We stress that the double-dual graviton field $C$ is determined {\it algebraically} from the Pauli-Fierz field $h$ up to $(2,2)$-diffeomorphisms.  

The equations (\ref{MR})-(\ref{Triviality2}) are the key equations of double-dualization.  In fact, the equation (\ref{Triviality2}) relating the potentials automatically implies by differentiation the double-dual relation (\ref{MR}) relating the curvatures, and so, one may view (\ref{Triviality2}) as the fundamental equation of double-dualization.

%In the gauge where $\xi_{\mu\nu \beta}=0$, this fundamental equation reduces to
%\be
%C_{\alpha \beta \mu \nu} =   - \frac{1}{2} \left(\eta_{\mu [\alpha} h_{\beta] \nu} - \eta_{\nu [\alpha} h_{\beta] \mu} \right) \, .\label{Triviality2Trivial}
%\ee
%When $\xi_{\mu\nu \beta}=0$,  the double-dual graviton is thus a mere ``$(2,2)$-dressing'' of the graviton, obtained by tensoring $h_{\mu \nu}$ with $\eta_{\alpha \beta}$ and making the appropriate projection. 

\subsubsection*{No exchange of equations of motion and Bianchi identities}

It follows from (\ref{MR}) that when the equations of motion for the Pauli-Fierz field hold ($R_{\mu \nu}=0$), the $(2,2)$ field $C$ is also on-shell, i.e. the double-trace $M_{\mu \nu}$ of the curvature $M_{\alpha_1 \alpha_2 \alpha_3 \beta_1 \beta_2 \beta_3}$ vanishes.  Conversely, if  $M_{\mu \nu} = 0$, then $R_{\mu \nu}=0$.

There is therefore no exchange of Bianchi identities with equations of motion when one goes to the double-dual graviton: equations of motion are mapped on equations of motion.

This is the reason why the relation between the graviton field $h_{\mu \nu}$ and it double-dual $C_{\alpha_1 \alpha_2  \beta_1 \beta_2}$ is {\it algebraic} (up to unavoidable gauge transformations) and simply given by (\ref{Triviality2}).  Double-duality can be defined algebraically directly in terms of the fields themselves, without having to go through the curvatures.  This is in sharp contrast with single dualization, which is algebraic in terms of curvatures but not so in terms of potentials, which are related by non-local expressions.

In terms of sources -- and although this question deserves further exploration --, there appears to be only sources of two types, electric and magnetic \cite{Hull:2001iu,Bunster:2013era}.  The source for the double dual graviton is the standard ``electric" energy-momentum tensor.  In 4 spacetime dimensions, dualization of the Schwarzschild solution gives Taub-NUT, and a further dualization of Taub-NUT on the second column brings back Schwarzschild.  There is no  new ``doubly magnetic" solution.

\subsubsection*{Cotton tensor}
%\label{subsec:Cottton}

As we already emphasized, the equation (\ref{Triviality2}) indicates that the double-dual graviton field is ``conformally flat'', i.e. the sum of a $(2,2)$-diffeomorphism parametrized by $\xi$ and a $(2,2)$-Weyl rescaling parametrized by $h$.  

One can understand this relation as follows. Although the Weyl tensor identically vanishes in 5 dimensions, not every $(2,2)$-field is conformally flat. What measures conformal flatness is the ``Cotton tensor'' $D_{\alpha \beta \lambda \mu}[C]$ of the $(2,2)$-field $C_{\alpha \beta \lambda \mu}$ defined in \cite{Henneaux:2016opm},
\be
 D_{\alpha \beta \lambda \mu}[C] = \frac{1}{3!} \epsilon_{\alpha \beta \rho \sigma \theta} \partial^\rho S^{\sigma \theta}_{\; \; \; \; \;  \lambda \mu}[C] ,
 \ee
 where $S_{\alpha \beta \lambda \mu}[C]$ is the ``Schouten tensor'' of $C$.   
 
 Now, in our case, $S_{\alpha \beta \lambda \mu}[C] =\frac{1}{18} R_{\alpha \beta \lambda \mu}[h]$  because of the Einstein equations for $h_{\mu \nu}$ (or equivalently, the double-tracefree condition on  $M_{\alpha_1 \alpha_2 \alpha_3 \beta_1 \beta_2 \beta_3}$), and so, the Cotton tensor $D_{\alpha \beta \lambda \mu}[C]$ is equal to zero on account of the Bianchi identity for $R_{\alpha \beta \lambda \mu}[h]$.  This implies that the double-dual of the graviton necessarily takes the conformally flat form (\ref{Triviality2})  \cite{Henneaux:2016opm} when it fulfills its equation of motion.
 
 In fact, as our derivation shows, one does not need the full equations of motion of the double graviton to derive this result, but only their weaker consequence  that the Cotton tensor $D_{\alpha \beta \lambda \mu}[C]$ vanishes.  This enables one to go off-shell while keeping the double-dual graviton field $C$ in the class of conformally flat fields, without assuming the stronger condition that the double-trace of its Riemann tensor is zero, or equivalently, that the graviton field $h$ is on-shell.  Indeed,  there is no need for $h_{\mu \nu}$ to fulfill the (linearized) Einstein equations for the $C$-field to exist and to be given by (\ref{Triviality2}). Vice-versa, there is no need for the double-dual graviton (taken in the class of conformally flat fields) to fulfill the double-trace condition $M_{\lambda \mu} = 0$ for the Pauli-Fierz field to exist.

\subsubsection*{Space of conformally flat $(2,2)$-tensors} 
\label{subsec:ConfTriv}

The equation (\ref{Triviality2}) defines a map from the space of the $\xi$'s and the $h$'s to the space of conformally flat $(2,2)$-tensors,
\be
(\xi_{\mu\nu \beta}, h_{\beta \nu}) \mapsto C_{\alpha \beta \mu \nu} = {\mathbb P}_{(2,2)}\left(\partial_{\alpha} \xi_{\mu\nu \beta} \right)  - \frac{1}{2} \left(\eta_{\mu [\alpha} h_{\beta] \nu} - \eta_{\nu [\alpha} h_{\beta] \mu} \right) \, .\label{Triviality22}
\ee 
Although surjective, this map is not injective, i.e. the parametrization of conformally flat $(2,2)$-tensors given by the pair $(\xi_{\mu\nu \beta}, h_{\beta \nu})$ involves redundancies.

The only ambiguity in $h_{\mu \nu}$ for a given $C_{\alpha \beta \mu \nu}$ is that it is determined up to a $(2,0)$-diffeomorphism $\partial_{(\mu} \zeta_{\nu)}$ since its curvature is completely determined. 
Defining the traceless part of $\xi_{\mu \nu \beta}$ and its trace $\xi_\mu$ as usual,
\be
\xi_\mu = \xi\indices{_{\mu\nu}^\nu}, \qquad  \wt{\xi}\indices{_{\mu\nu}^\rho} = \xi\indices{_{\mu\nu}^\rho} + \frac{1}{2} \delta^\rho_{[\mu} \xi_{\nu]}\, ,
\ee
one sees that the $(2,0)$-diffeomorphism $h_{\mu \nu} \rightarrow h_{\mu \nu} + \partial_{(\mu} \zeta_{\nu)}$ can be compensated by the shift $\xi_\mu \rightarrow \xi_\mu + 2 \zeta_\mu$ of the trace of $\xi_{\mu \nu \beta}$:  the combined transformations leave indeed $C_{\alpha \beta \mu \nu}$ invariant. 

But this is not the only redundancy in the parametrization of conformally flat tensors, because ${\mathbb P}_{(2,2)}\left(\partial_{\alpha} \xi_{\mu\nu \beta} \right)$ can be traceful even if $\xi_{\mu \nu \beta}$ itself is traceless.  The shift $\wt{\xi}_{\mu\nu\rho} \rightarrow \wt{\xi}_{\mu\nu\rho} + \wt{\Lambda}_{\mu\nu\rho}$ with
$$
\wt{\Lambda}_{\mu\nu\rho} = \pd_{[\mu} A_{\nu]\rho} - \pd_\rho A_{\mu\nu} - \frac{3}{4} \eta_{\rho[\mu} \pd^\lambda A_{\nu]\lambda}
$$
($A_{\mu \nu}$ is an arbitrary antisymmetric tensor), combined with the diffeomorphism $h_{\mu \nu} \rightarrow h_{\mu \nu} + \partial_{(\mu} (-\frac12\partial^\lambda A_{\nu) \lambda})$ leaves also $C_{\alpha \beta \mu \nu}$ invariant. 

The fact that the parametrization of conformally flat $(2,2)$-tensors provided by the pair $(\xi_{\mu\nu \beta}, h_{\beta \nu})$ is redundant is not a problem for the subsequent discussion and we shall therefore not attempt to ``gauge-fix'' it.

It is at this point useful to decompose the tensor $C_{\alpha \beta \mu \nu}$ into a traceless part and a traceful part.  One has
\be
\wt{C}\indices{^{\mu\nu}_{\rho\sigma}} = C\indices{^{\mu\nu}_{\rho\sigma}} - \frac{4}{3} \delta^{[\mu}_{[\rho} C\indices{^{\nu]} _{\sigma]}} + \frac{1}{6} \delta^{\mu\nu}_{\rho\sigma} C\,  \quad C\indices{^\mu_\nu} = C\indices{^{\mu\rho}_{\nu\rho}}, \quad C = C\indices{^{\mu\nu}_{\mu\nu}}.
\ee
The traceless component $\wt{C}$  of the double dual graviton does not involve $h_{\mu \nu}$ and depends on $\wt{\xi}$ only. It is therefore pure gauge.
%One finds explicitly
%\begin{align}
%\wt{C}\indices{^{\mu\nu}_{\rho\sigma}} &= \wt{\mathbb P}_{(2,2)}\left(\partial_{\rho} \wt{\xi}\indices{^{\mu\nu}_{\sigma}} \right) \\
%&= \frac{1}{2} \left( \partial_{[\rho} \wt{\xi}\indices{^{\mu\nu}_{\sigma]}} + \partial^{[\mu} \wt{\xi}\indices{_{\rho\sigma}^{\nu]}} \right) - \frac{1}{3} \delta^{[\mu}_{[\rho} \left( \pd \cdot \wt{\xi}\indices{_{\sigma]}^{\nu]}} + \pd \cdot\wt{\xi}\indices{^{\nu]}_{\sigma]}} \right), \qquad \left(\pd \cdot \wt{\xi}_{\mu\nu} \equiv \pd^\rho \wt{\xi}_{\rho\mu\nu} \right)
%\end{align}
%and $\wt{C}\indices{^{\mu\nu}_{\rho\sigma}} $ is therefore pure gauge.

One can trade $h_{\mu \nu}$ for the trace $C_{\mu\nu}$ in the parametrization of  $C_{\alpha \beta \mu \nu}$.  Taking traces of \eqref{Triviality22}, one easily gets 
\be
\frac{3}{4}\, h_{\alpha \lambda}   = - C_{\alpha\lambda} + \frac{1}{8} \eta_{\alpha\lambda} C + \frac{1}{2} \pd_{(\alpha} \xi_{\lambda)} + \frac{1}{2} \pd^\mu \xi_{\mu(\alpha \lambda)} - \frac{1}{8} \eta_{\alpha\lambda} \pd^\mu \xi_{\mu} \, .\label{hC}
\ee
The change of parametrization $(\xi_{\mu\nu \beta}, h_{\beta \nu}) \leftrightarrow (\xi_{\mu\nu \beta}, C_{\beta \nu})$ is clearly invertible. 

The equation (\ref{hC}) indicates that $h_{\lambda \mu}$ is not  determined only by the trace of the $(2,2)$-tensor, even up to a linearized $(2,0)$-diffeomorphism.  There are additional contributions coming from $\xi_{\mu\nu \beta}$ because ${\mathbb P}_{(2,2)}\left(\partial_{\alpha} \xi_{\mu\nu \beta} \right)$ has in general a non-zero trace.  

The need for the presence of such contributions can be understood from the fact that  if one shifts $\xi_{\mu\nu\beta}$ as
\be
\xi_{\mu\nu\beta} \rightarrow \xi_{\mu\nu\beta} + \Lambda_{\mu\nu\beta}
\ee
the traceful part of the tensor $C_{\alpha \beta \mu \nu}$ will remain invariant if $C_{\alpha \lambda}$ is transformed at the same time as
\begin{eqnarray}
&&C_{\alpha \lambda} \rightarrow C_{\alpha \lambda} + \delta C_{\alpha \lambda} , \\
&&\delta C_{\alpha \lambda} = \frac12 \pd_{(\alpha} \Lambda_{\lambda)}  +  \frac{1}{2} \pd^\mu \Lambda_{\mu(\alpha \lambda)},  \label{eq:TransTrace}
\end{eqnarray}
an expression that is not the symmetrized derivative of a vector, i.e. which does not take the form of the transformation  of the graviton.   Thus the relationship between $h_{\mu \nu}$ and the trace $C_{\mu\nu}$ must involve compensating terms in $\xi$ that account for these different behaviours.

\subsection{Lagrangian for the $C$-field}
\label{sec:LagrangianC}

The previous considerations naturally suggest a Lagrangian for the double-dual of the graviton. 

The standard Lagrangian for a $(2,2)$ field constructed along the lines of \cite{Curtright:1980yk,Aulakh:1986cb,Labastida:1986gy,Labastida:1987kw} and explicitly written down in e.g. \cite{Burdik:2000kj,Bizdadea:2003ht,Boulanger:2004rx} is not the correct Lagrangian for the double-dual graviton because the equations that follow from it imply that the single trace of the double-dual curvature $M_{\alpha_1 \alpha_2 \alpha_3 \beta_1 \beta_2 \beta_3}$ vanishes, and hence, as we explained, that $M_{\alpha_1 \alpha_2 \alpha_3 \beta_1 \beta_2 \beta_3}$ itself vanishes since $M_{\alpha_1 \alpha_2 \alpha_3 \beta_1 \beta_2 \beta_3}$ has no Weyl part.  The standard $(2,2)$-theory has no degree of freedom in $5$ dimensions, in sharp contrast to the standard $(2,1)$-theory that describes the (single) dual graviton.

The correct equations of motion for the double dual graviton are that the double trace of the curvature $M_{\alpha_1 \alpha_2 \alpha_3 \beta_1 \beta_2 \beta_3}$ is zero, as we recalled.  As we now show, these equations can easily be derived from a variational principle, which is a direct rewriting of the Pauli-Fierz variational principle.

One can view the Pauli-Fierz action $S[h]$ as a functional of both $h_{\mu \nu}$ and $\xi_{\mu \nu \rho}$ that depends trivially on $\xi_{\mu \nu \rho}$, namely, it is constant under any change of $\xi_{\mu \nu \rho}$, ${\delta S}/{\delta \xi_{\mu \nu \rho}} \equiv 0$.  The theory defined by the action $S[h,\xi] \equiv S[h]$ is clearly equivalent to the original Pauli-Fierz theory, since the variational equations 
\begin{equation}
\frac{\delta S}{\delta h_{\mu \nu}} = 0  \label{eq:VarEq5A}
\end{equation}
are unchanged, while the equations
\begin{equation}
\frac{\delta S}{\delta \xi_{\mu \nu \rho}} = 0 \label{eq:VarEq5B}
\end{equation}
are of the form $0=0$ (empty), which tells us that the field $\xi_{\mu \nu \rho}$ is pure gauge -- in fact invariant under the shift symmetry $\xi_{\mu \nu \rho} \rightarrow \xi_{\mu \nu \rho} + \Lambda_{\mu \nu \rho}$, where the gauge parameters $\Lambda_{\mu \nu \rho}$, which have the $(2,1)$ Young symmetry,  are arbitrary.

One can equivalently view the action $S[h,\xi]$ as a functional defined in the space of conformally flat $(2,2)$-tensors.  It depends indeed only on $C_{\alpha \beta \mu \nu}$ (assumed to be conformally flat, i.e. with zero Cotton tensor) and not on the specific choice of $\xi_{\alpha \beta \mu}$ and $h_{\mu \nu}$ entering its decomposition,  since $S[h,\xi]$ involves only $h_{\mu \nu}$, which is determined by the conformally flat $(2,2)$-tensor up to a diffeomorphism under which the action is invariant.

We have thus derived a variational principle for the $C$-field, assumed to be in the class of conformally flat $(2,2)$-tensors.

Having to impose the conformal flatness condition by hand is a bit awkward. One would like to free oneself from this constraint and formulate a variational principle in which arbitrary $(2,2)$-tensors can be considered. 

One way to achieve this goal relies on the following observation.
If one injects in the Pauli-Fierz action the expression (\ref{hC}) of $h_{\mu \nu}$ in terms of $C_{\mu \nu}$ and $\xi_{\alpha \beta \mu}$, one gets an action $S[C_{\alpha \beta \mu \nu}, \xi_{\alpha \beta \mu}]$,
\begin{equation}
S[C_{\alpha \beta \mu \nu }, \xi_{\alpha \beta \mu}] = S^{\textrm{Pauli-Fierz}}\left[h_{\alpha \lambda} = H_{\alpha \lambda}(C, \partial \xi  ) \right] \label{eq:FinalAction}
\end{equation}
 which yields by construction the correct equations of motion.  Here, $H_{\alpha \lambda}(C, \partial \xi  )$ is the expression of $h_{\alpha \lambda}$ in terms of $C$ and  $\partial \xi$ following from (\ref{hC}).  Varying the action (\ref{eq:FinalAction}) with respect to the trace $C_{\mu \nu}$ and $\xi_{\alpha \beta \mu}$  gives equations equivalent to  (\ref{eq:VarEq5A}) and (\ref{eq:VarEq5B}), i.e. the Pauli-Fierz equations for $H_{\alpha \lambda}$,  since the new action is obtained from the old one by a mere (invertible) change of variables.   Varying the action with respect to the remaining traceless components $\wt{C}\indices{^{\mu\nu}_{\rho\sigma}}$ gives nothing ($0 = 0$), since these components do not appear in the action.  The field $C\indices{^{\mu\nu}_{\rho\sigma}}$ is unconstrained in the variational principle based on (\ref{eq:FinalAction}).

The action $S[C_{\alpha \beta \mu \nu }, \xi_{\alpha \beta \mu}] $ so constructed possesses a huge gauge invariance:
\begin{itemize}
 \item Because only the trace of $C_{\alpha \beta \lambda \mu}$ appears, the trace-free part of $C_{\alpha \beta \lambda \mu}$  enjoys a ``shift symmetry" that accounts for its non-appearance in the action,
\begin{equation}
\delta C_{\alpha \beta \lambda \mu} = \wt{\Omega}_{\alpha \beta \lambda \mu} , \qquad \delta \xi_{\alpha \beta \mu} = 0 \end{equation}
where $ \wt{\Omega}_{\alpha \beta \lambda \mu} $ is an arbitrary trace-free gauge parameter with $(2,2)$ Young symmetry.  This can be equivalently written $\delta \wt{C}_{\alpha \beta \lambda \mu} = \wt{\Omega}_{\alpha \beta \lambda \mu} $, $\delta C_{\mu \nu} = 0$, $\delta \xi_{\alpha \beta \mu} = 0$.

\item Similarly, there is a shift symmetry for $\xi_{\alpha \beta \mu}$ under which the trace $C_{\mu \nu}$ transforms as in (\ref{eq:TransTrace}),
\be
\delta \xi_{\mu\nu\beta}= \Lambda_{\mu\nu\beta}, \qquad \delta C_{\alpha \lambda} = \frac12 \pd_{(\alpha} \Lambda_{\lambda)}  +  \frac{1}{2} \pd^\mu \Lambda_{\mu(\alpha \lambda)} .
\ee
For the variation  $\delta \wt{C}_{\alpha \beta \lambda \mu}$ of the traceless part of the $(2,2)$-tensor under those transformations, one can take anything.  One possibility is simply to take $\delta \wt{C}_{\alpha \beta \lambda \mu} = 0$.  Another possibility is to take $\delta \wt{C}_{\alpha \beta \lambda \mu}=  \wt {\mathbb P}_{(2,2)}\left(\partial_{\alpha} \Lambda_{\mu\nu \beta} \right)$ so that $\delta C_{\alpha \beta \lambda \mu}=  \mathbb P_{(2,2)}\left(\partial_{\alpha} \Lambda_{\mu\nu \beta} \right)$.  The invariance of the action under $(2,2)$-diffeomorphisms of the $C$-field might seem surprising at first because the action does not depend only on the corresponding curvature invariants.  But there is no contradiction because the action also involves the compensating field $\xi_{\alpha \beta \mu}$ in precisely the right (compensating!) way.

\item Finally, the trace of $C_{\alpha \beta \lambda \mu}$ itself enjoys the gauge symmetry inherited from the linearized diffeomorphism invariance of the Pauli-Fierz action from which (\ref{eq:FinalAction}) comes.  This gauge symmetry can be written as
\begin{equation}
\delta C_{\alpha \lambda} = -\frac32 \pd_{(\alpha} \zeta_{\lambda)} - \frac12 \eta_{\alpha \lambda} \partial^\mu \zeta_\mu , \quad \delta \wt{C}_{\alpha \beta \lambda \mu} = 0, \quad \delta \xi_{\mu\nu\beta} = 0.
\end{equation}
\end{itemize}

The action $S[C, \xi]$ given by (\ref{eq:FinalAction}) is the central result of this section.  It depends on the dual graviton field $C_{\alpha \beta \lambda \mu}$ and on the additional field $ \xi_{\lambda \mu \beta}$. This additional field can be gauged away using its shift gauge symmetry, as can the traceless part of $C_{\alpha \beta \lambda \mu}$.  When the gauge conditions $\xi_{\lambda \mu \beta}= 0$ and $\wt{C}_{\alpha \beta \lambda \mu}= 0$ are imposed, the action simply reduces to the Pauli-Fierz action.

It would be worthwhile to compare our work with the action proposed in the interesting articles \cite{Boulanger:2012df,Boulanger:2012mq}, where a very different approach is followed to derive an action for the double dual graviton field. The authors of \cite{Boulanger:2012df,Boulanger:2012mq} extend to the double dual case the parent action method used in \cite{Boulanger:2003vs} for first-dualization of the graviton, resulting in an action different from ours and containing, as here, (different) additional fields and (different) additional gauge symmetries (see also \cite{Chatzistavrakidis:2019len} for recent work on this type of actions).

\subsection{Dimensional reduction of a $(2,2)$-field from 6 to 5 spacetime dimensions}
Our considerations are relevant in the dimensional reduction of a $(2,2)$-field from 6 to 5 spacetime dimensions. A $(2,2)$-field in 6 dimensions is not equivalent to the Pauli-Fierz field.  Upon dualization, one gets another $(2,2)$-field, 
\be
^*\left(\yng(2,2)\right) \sim \yng(2,2)
\ee
which is independent from the original $(2,2)$-field unless one imposes a self-duality condition as in \cite{Hull:2000ih,Hull:2000zn,Hull:2000rr} (see \cite{Henneaux:2016opm} for the implementation of the self-duality condition in the action, and \cite{Lekeu:2018kul} for further considerations on dimensional reduction).  A general (not self-dual) massless gauge $(2,2)$ field is described by a Curtright-like action (see \cite{Bizdadea:2003ht,Boulanger:2004rx}) and has $10$ physical degrees of freedom in six spacetime dimensions.  Upon dimensional reduction to 5 dimensions, it decomposes as
\be
\yng(2,2) \; \; \; \rightarrow \; \; \; \yng(2,2) \oplus \yng(2,1) \oplus \yng(2)
\ee
where each five-dimensional field on the right-hand side  of this equation is described by the relevant Curtright-like action (which coincides of course with the Pauli-Fierz action for the $\yng(2)$-field).  

Now, in 5 dimensions, a $(2,2)$ field described by a Curthright-like action (which is {\em not} the action of the previous subsection) has no physical degree of freedom, while, as we have seen, the $(2,1)$-field is dual to the Pauli-Fierz field. Thus, the dimensional reduction of a general (not self-dual) massless $(2,2)$-field gives two Pauli-Fierz fields, each carrying 5 degrees of freedom (the number of helicity states of a massless spin-$2$ particle). A self-duality condition in 6 dimensions \cite{Hull:2000zn,Hull:2000rr} equates these two fields.

The counting is very similar to the counting relevant to the dimensional reduction from 4 to 3 dimensions of the Pauli-Fierz field.  One has
\be 
\yng(2) \; \; \; \rightarrow \; \; \; \yng(2) \oplus \yng(1) \oplus \bullet
\ee
where $\bullet$ stands for the scalar representation.  In 3 dimensions, gravity carries no local degrees of freedom and so the first term $\yng(2)$ does not contribute to the number of local degrees of freedom, in much the same way as the representation $\yng(2,2)$ did not contribute above.  Furthermore, a vector is dual to a scalar.  The 3-dimensional version of the theory contains thus two scalars (one cannot impose a self-duality condition in 4 dimensions (with Lorentz signature) that would equate those real scalars).

\section{Extension to higher dimensions}

Similar results hold in all dimensions $D \geq 4$.  In fact, already in $D=4$ dimensions one can see that the double dual graviton is somewhat trivial.  The single dual graviton  is a symmetric tensor $b_{\mu \nu}$ not locally related to the Pauli-Fierz field $h_{\mu \nu}$ \cite{Casini:2003kf,Boulanger:2003vs}.  The double-dual graviton is also a symmetric tensor $C_{\mu \nu}$ but it is however locally related to $h_{\mu \nu}$.  Indeed, its curvature is (on-shell) equal to the curvature of $h_{\mu \nu}$ and so $C_{\mu \nu} = h_{\mu \nu}$ up to linearized diffeomorphisms.

In higher dimensions, the double graviton is a $(D-3,D-3)$-tensor.  Its curvature is a $(D-2,D-2)$-tensor.  Now, a $(D-2,D-2)$-tensor is completely determined by its $(D-4)$-th trace, which is a $(2,2)$-tensor.  By the duality conditions, this $(2,2)$-tensor is (on-shell and up to a numerical factor) equal to the Riemann tensor of the Pauli-Fierz field. Therefore, the double graviton field is an appropriately symmetrized product of $(D-4)$ $\eta_{\lambda \mu}$'s with the Pauli-Fierz field (up to a gauge transformation), something that can be checked directly by comparing the expressions of the double dual of the curvature in terms of $\partial^2h$ and $\partial^2 C$. 
So again, the double dual graviton is an algebraic rewriting of the original Pauli-Fierz field.

\section{Comments and Conclusions}
\label{Conclusions}

The fact that the double dual graviton plays a more minor role is somewhat disappointing, since it does not allow a full triality where the double dual graviton would be on the same footing as the graviton and its dual.   This might explain why the double dual graviton does not appear in the spectrum of the $E_{10}$ or $E_{11}$ (conjectured) reformulations of maximal supergravity/M-theory, where there is no $(8,8)$ field (only fields characterized by a Young tableau with $3k$ boxes can appear). [Different, interesting ``duals" are considered in \cite{Riccioni:2006az} by adjoining columns of $D-2$ boxes to the Young tableaux, but these  do not correspond to the duals defined here through dualizations of the curvature.]

We close by noting that  in the Hamiltonian formulation, ``duality conjugate" is equivalent to ``canonically conjugate" (see e.g. \cite{Bunster:2013oaa}). Our analysis matches the property that canonical variables come in conjugate pairs (and not in conjugate triplets, or  even n-plets).  
This is true for any higher spin gauge field and not just spin 2.  And indeed, the Hamiltonian analysis yields two independent conjugate prepotentials tied to duality, one associated with the original field and one associated with its single dual, without room for a third prepotential (or more) \cite{Henneaux:2015cda,Henneaux:2016zlu}. From the point of view of the Hamiltonian description, higher spin gauge fields are similar to spins one and two and admit two independent dual descriptions only.

That higher spin gauge fields admit only two independent dual descriptions when duality is defined through Hodge duality of the curvature tensor can in fact also be seen by direct algebraic manipulations.  Take for instance a spin-$3$ gauge field in 5 dimensions, described by a symmetric tensor field $h_{\mu \nu \rho} =  h_{(\mu \nu \rho)}$, corresponding to the Young tableau $\yng(3)$.  The curvature is a $\yng(3,3)$-tensor.  The equations of motion can be taken to be that the trace of the curvature tensor is zero \cite{Bekaert:2003az}. 

The first dual is a $\yng(3,1)$-tensor, with curvature in the representation 
$$
\yng(3,3,1).
$$
The double dual is a $\yng(3,2)$ with curvature in the representation
$$
\yng(3,3,2).
$$
The same reasoning as for the graviton shows that the double dual curvature is of the form of an appropriately symmetrized product of $\eta_{\alpha \beta}$ with the original $\yng(3,3)$-curvature, and that the double dual spin-$3$ field is  of the form of an appropriately symmetrized product of $\eta_{\alpha \beta}$ with the original spin-$3$ field $h_{\mu \nu \rho}$ and so is not an algebraically independent object.  Similar equations relate the triple dual field to the single dual one.

The considerations in this paper are valid for the linear theory and do not preclude nonlinear surprises.

%%%%%%%%%%%%%%%%%%%%%%%%%%%%%%%%%%%%%%%%%%%%%%%%%%%%%
\section*{Acknowledgments} 
 We thank Nicolas Boulanger and Chris Hull for useful discussions. V.L. is Research Fellow at the Belgian F.R.S.-FNRS, and would like to thank Imperial College London for hospitality during the course of this work. This work was partially supported by the ERC through the ``High-Spin-Grav" Advanced Grant and under the European Union Horizon 2020 research and innovation programme ("Exceptional Quantum Gravity", grant agreement No 740209), and by FNRS-Belgium (convention FRFC PDR T.1025.14 and convention IISN 4.4503.15).
 
% \break

%\bibliographystyle{utphys}
%\bibliography{bibliography}

\providecommand{\href}[2]{#2}\begingroup\raggedright\endgroup

\end{document}